\documentclass[pra,twocolumn,showpacs,preprintnumbers,amsmath,amssymb,superscriptaddress, longbibliography]{revtex4-1}
\usepackage{graphicx}
\usepackage{subfigure}
\usepackage{dcolumn}
\usepackage{bm}

\begin{document}


\title{Flux unwinding in the lattice Schwinger model}

\author{Chris Nagele}
\affiliation{New York University Shanghai, 1555
Century Ave, Pudong, Shanghai 200122, China}
\affiliation{Center for Cosmology and Particle Physics, New York University,
New York, 10003, USA}

\author{J. Eduardo Cejudo}
\affiliation{New York University Shanghai, 1555 Century Ave, Pudong, Shanghai 200122, China}

\author{Tim Byrnes}
\affiliation{New York University Shanghai, 1555
Century Ave, Pudong, Shanghai 200122, China}
\affiliation{State Key Laboratory of Precision Spectroscopy, School of Physical and Material Sciences, East China Normal University, Shanghai 200062, China}
\affiliation{NYU-ECNU Institute of Physics at NYU Shanghai,
3663 Zhongshan Road North, Shanghai 200062, China}
\affiliation{National Institute of Informatics, 2-1-2 Hitotsubashi, Chiyoda-ku, Tokyo 101-8430, Japan}
\affiliation{Department of Physics, New York University, New York, NY 10003, USA}

\author{Matthew Kleban}
\affiliation{Center for Cosmology and Particle Physics, New York University,
New York, 10003, USA}
\affiliation{Department of Physics, New York University, New York, NY 10003, USA}


\date{\today}

\begin{abstract}

We study the dynamics of the massive Schwinger model on a lattice using exact diagonalization.  When periodic boundary conditions are imposed, analytic arguments indicate that a non-zero  electric flux in the initial state can ``unwind" and decrease to a minimum value equal to minus its initial value, due to the effects of a pair of charges that repeatedly traverse the spatial circle.  Our numerical results support the existence of this flux unwinding phenomenon, both for initial states containing a charged pair inserted by hand, and when the charges are produced by Schwinger pair production.  We also study  boundary conditions where charges are confined to an interval and flux unwinding cannot occur, and the massless limit, where our results agree with the predictions of the bosonized description of the Schwinger model.   \end{abstract}


\maketitle

\section{Introduction}

The massive Schwinger model \cite{schwinger1962gauge} --- quantum electrodynamics in one space and one time dimension --- is a fascinating quantum field theory that has been studied intensively since the 1950s.  It has a wide set of applications: as a simple example of a quantum gauge theory, as an Abelian theory that nevertheless exhibits a linearly growing potential between charges and hence a kind of confinement, as a theory that exhibits a prototype strong/weak duality via bosonization,  and even to models of cosmic inflation in string theory \cite{COLEMAN1975267, Coleman:1976uz, damico2012}.

Most work  on the Schwinger model has focused on its static properties, such as its spectrum of excitations, the value of the chiral condensate, etc.  There has been relatively little work, either analytical or numerical, on time-dependent phenomena in the Schwinger model.  Two recent works include Hebenstreit et.~al., who considered the dynamics of string breaking in the massive Schwinger model using a numerical technique where the gauge field is treated classically/statistically \cite{hebenstreit2013}, and Buyens et.~al.~who studied real-time evolution of the wavefunction 
using the Matrix Product States formalism in the thermodynamic limit \cite{buyens2016}. 

\begin{figure}[t]
\centering     
\includegraphics[width=\columnwidth]{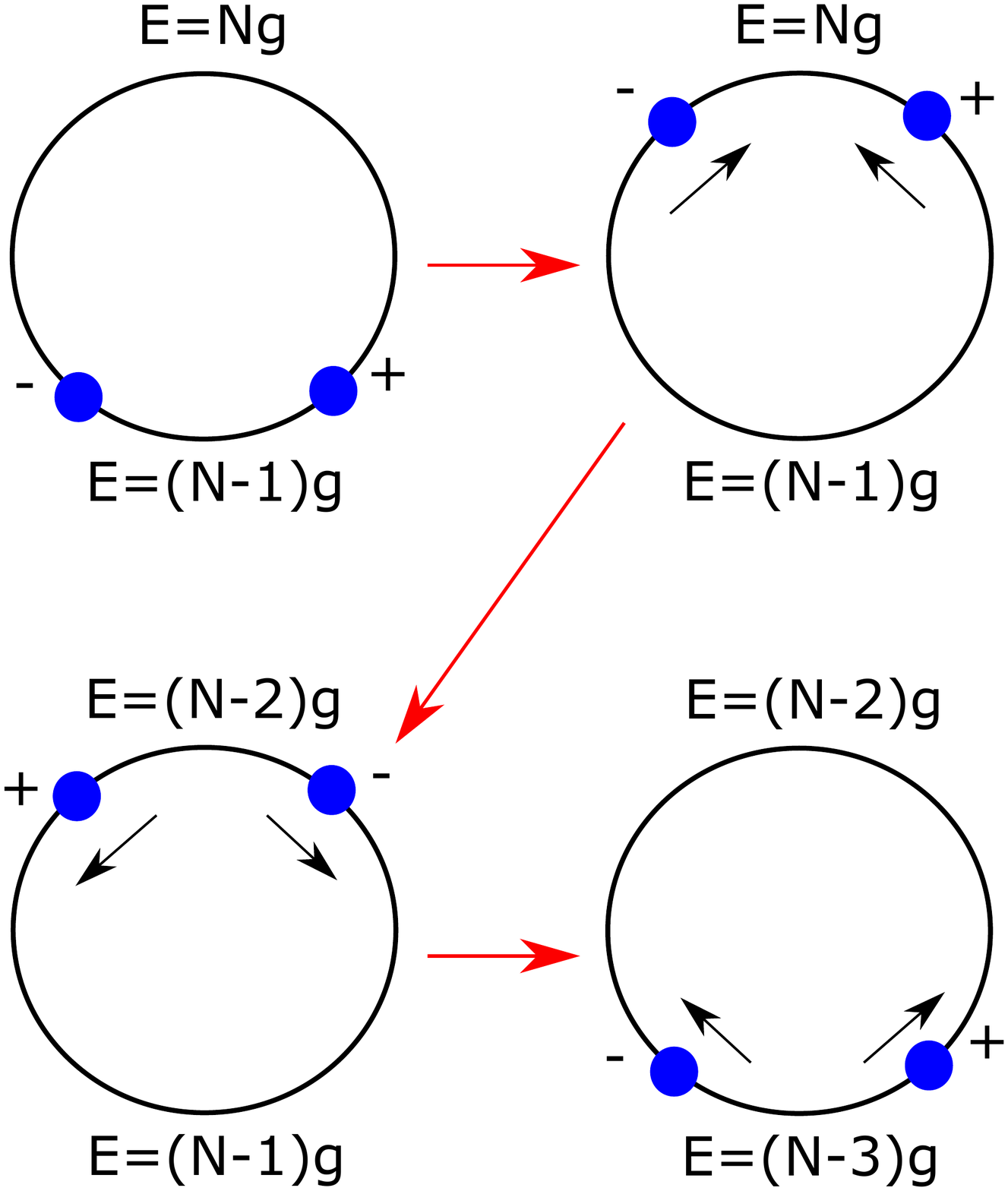}
\caption{Schematic representation of the dynamics of a pair particle-antiparticle under 
an external electric field (quench) $\alpha$.
The electric field accelerates the fermions  in opposite directions, lowering the average electric field on each circuit.}
\label{figloop}
\end{figure}

Despite the absence of electromagnetic waves in one spatial dimension, the electric field in the Schwinger model is generally time-dependent because charged particles move and affect its value.  These particles  can be spontaneously produced by Schwinger pair production in the quantum theory \cite{heisenberg1935}, or simply be present in the initial state.  
In Ref.~\cite{kleban2011}, a new time-dependent phenomenon was discovered in the Schwinger model (and a broad class of other theories) with spatially periodic boundary conditions. 
A solution to the classical theory with no charges is a homogeneous, time-independent electric field that winds around the spatial circle.  If a  pair of equal and opposite charges is present, the field accelerates the charges in opposite directions until they collide at some point on the opposite side of the circle (see Fig.~\ref{figloop}).  If the charges transmit through each other, they will continue in the same direction, unwinding two units of charge 
on each circuit (charge and field strength have the same units in one dimension).   As a result, the initial value of the field will steadily decrease.  In the absence of any other dynamics, the momentum of the charges causes the electric field to overshoot zero, decreasing to a value with equal magnitude and opposite sign as the initial field. 
  This is sharply in contrast with the case of the infinite line or  boundary conditions on an interval that forbid charges from crossing, where a single charged pair can at most reduce the field by two units.

This mechanism is known as a flux discharge cascade or flux unwinding \cite{kleban2011}, and is related to the phenomenon of  ``axion monodromy'' \cite{2008PhRvD..78j6003S, 2010PhRvD..82d6003M}.  Note that the unwinding mechanism depends crucially on the ability of an electron and positron to transmit directly through each other without reflecting, annihilating, or forming a bound state.  If any of these other processes occur with non-negligible probability, unwinding may still happen some of the time or in one branch of the wavefunction, but it will not necessarily be the dominant process.

Generalizations of this unwinding process are potentially of  interest to the theory of cosmic inflation \cite{damico2012, 2013PhLB..725..218D}.  In  theories such as string theory and supergravity with higher-dimensional charged objects (branes) and the higher-form analogs of electromagnetic fields they couple to, the gravitational effect of the energy in the field can drive exponential or quasi-exponential expansion of space.    During the unwinding process the energy gradually decreases, so that the rate of this slow-roll inflationary expansion reduces gradually and then comes to an end.  Furthermore the initial state prior to (the analog of) Schwinger pair production rapidly inflates and produces an exponentially large volume, and therefore arguably constitutes a natural initial condition for the universe.

In this paper, we examine the lattice version of the Schwinger model and study several time-dependent phenomena in a variety of parameter regimes and for several different initial states. Most prior  numerical work on the lattice Schwinger model was restricted to  what is referred to in the literature as ``open boundary conditions" (OBC), where the electric field is fixed at the edges and charges reflect off the boundaries \cite{banuls2013matrix,banuls2015thermal,banuls2017density}.  With periodic boundary conditions (PBC)  the theory has an extra quantum mechanical degree of freedom, which can be thought of as the electric field at one lattice site \cite{manton1985}.  For a fixed number of lattice sites we exactly diagonalize the full Hamiltonian, and establish that flux unwinding  indeed occurs in the non-perturbative lattice theory with PBC when a massive charged pair is inserted in the initial state.  We  observe that the electrons and positrons can transmit through each other in this regime with a fairly high probability.  We also study the dynamics of the  model with OBC.  Flux unwinding cannot occur with OBC, but our simulations clearly show that positive and negative charges can transmit through each other with high probability.  Finally, we study the time-evolution of the zero-electric field ground state and show that Schwinger pair production occurs and leads to flux unwinding.

This paper is structured as follows.  In Section \ref{sec:schwinger} we review the discrete version of the Schwinger model. In Section \ref{sec:exact} we describe our numerical techniques,  the initial states we will consider, and the observables we will compute.  In Section \ref{sec:results} we present our results, and in Section \ref{sec:conc} we conclude.

\section{The Schwinger Model on a Lattice}
\label{sec:schwinger}

The Hamiltonian for the continuum Schwinger model is that of Quantum Electrodynamics (QED) in one spatial dimension \cite{schwinger1962gauge,lowenstein1971quantum,byrnes2002density} 
\begin{align}
{\cal H} & = \int dy \Big[ -i \bar{\psi} \gamma^1 ( \frac{d}{dy} + i g A_1) \psi 
 + m 
\bar{\psi} \psi + \frac{E^2}{2} \Big].
\label{continuumham}
\end{align}
We work in natural units with $c = \hbar = 1$.  Here $ E $ is the electric field operator, the vector potential $A_1 $ is related to the electric field by $ E= -\frac{d A_1}{dt} $ because  we choose the gauge $ A_0 = 0 $, $ \psi $ is the two component field operator for the electrons and positrons, $ \bar{\psi} = \psi^\dagger \gamma_0 $, $ m $ is the mass of the electron, $ g $ is the charge. The $ \gamma^\mu $ matrices of dimension $ 2 \times 2 $ are defined by $\{ \gamma^\mu, \gamma^\nu \} = 2 \eta^{\mu \nu}$
where $ \eta^{\mu \nu} $ is the Minkowski metric with diagonal elements of $ (-1,1) $ and zero otherwise. The indices $ \mu, \nu $ run from 0 to 1.  Note that the charge $g$ has dimensions of mass, as does the electric field $E$.

In one spatial dimension,  Gauss' law takes the form 
\begin{align}
E(y) = F+ g \int_0^y dy' \, j^0 (y') 
\label{Gausscont}
\end{align}
where $ j^0   = \psi^\dagger \psi $ and $ F $ is a constant background electric field at the position $ y = 0 $.  The nature of Gauss' law in one dimension is that the electric field is constant when the charge density is zero, and changes by $g $ across the position of a charge $ g $.  

 This gives rise to a confining interaction between the electrons and positrons, as the constant field between them corresponds to a linearly growing potential.  
Due to the lack of electromagnetic waves in one dimension, there are no local degrees of freedom associated to the electric field, except possibly for the single global degree of freedom $F = E(0)$ (depending on the boundary conditions). 

The Hamiltonian (\ref{continuumham}) involves both the degrees of freedom for the particles (electrons and positrons) and the electric field. In order to numerically compute the Hamiltonian, the standard procedure
is to perform a  discretization by considering an staggered
lattice, where electrons and positrons occupy odd and even sites respectively \cite{kogut1975hamiltonian,carroll1976lattice}.  Starting from the discrete fermionic Hamiltonian, we apply a Jordan-Wigner transformation in order to map the fermion operators to spins \cite{banks1976strong}. The Hamiltonian in this formulation is
\begin{align}
H = & x \sum_{n=1}^N \Big[ \sigma^+_n  e^{i \theta_n} \sigma^-_{n+1} + \sigma^-_n e^{-i \theta_n} \sigma^+_{n+1} \Big] \nonumber \\
& + \sum_{n=1}^N \Big[ ( L(n) + \alpha)^2 + \frac{\mu}{2} ( 1+ (-1)^n \sigma^z_n) \Big]
\label{latticeham}
\end{align}
where $ \sigma^{x,y,z} $ are Pauli matrices and $ \sigma^\pm = (\sigma^x \pm i \sigma^y)/2 $.  The parameters are defined by
$$ 
\mu \equiv  \frac{2m}{g^2 a}, \,\,\,\, x \equiv \frac{1}{g^2a^2}, \,\,\,\, \alpha \equiv \frac{F}{g},
$$
where $a$ is the distance between lattice sites.  The dimensions have been scaled out of the Hamiltonian $H$ in (\ref{latticeham}); it is related   to the continuum Hamiltonian $ {\cal H}$ by  $ {\cal H} =  \lim_{a\rightarrow 0} \frac{ag^2}{2} H $.  
Due to the staggered lattice, spin up at an even site represents
a positron and spin down represents the vacuum;  spin down at an odd site 
represents an electron and spin up represents the vacuum. The number of lattice sites $ N $ is always assumed to be even.  In the spin language, states of zero charge correspond to states which have a total magnetization $ \sum_n \sigma^z_n = 0 $.

In \eqref{latticeham} we introduced the lattice electric field operator $ L(n) $, with eigenstates
\begin{align}
L(n) | l \rangle = l | l \rangle
\label{leigenstates}
\end{align}
where $ l $ is an integer.  It is conjugate to the vector potential according to $ A_1(x) \leftrightarrow - \frac{\theta(n)}{ag} $ and obeys the canonical commutation relation $ [ \theta(n), L(m)] = i \delta_{nm} $.  The exponential of the vector potential acts as a shift operator for  the eigenstates of $ L $:
\begin{align}
e^{\pm i \theta} | l \rangle = | l \pm 1 \rangle  .
\label{lhopping}
\end{align}
Gauss' law on the lattice is
\begin{align}
\label{Gauss}
L(n) - L(n-1) = \frac{1}{2}( \sigma^z_n + (-1)^n).  
\end{align}
This means that in any eigenstate of  spin and given the value of the electric field at the boundary $ L(0) + \alpha $, the electric field at every other point is determined everywhere, and changes only by $\pm1$ or $0$ from one site to the next. 

For the case of OBC we will set the electric field at the boundary to 
\begin{align}
E(0)/g=L(0) + \alpha = \alpha,
\end{align}
so that $L(0) = 0$.  Since $L(n)$ is then determined by \eqref{Gauss}, for OBC the electric field is not a quantum mechanical degree of freedom.  Instead, there  is a continuous family of OBC Hamiltonians indexed by the parameter $ \alpha $. Eliminating the electric field degree of freedom we can write the lattice Hamiltonian for OBC as
\begin{align}
H_{\text{OBC}}  & =  x\sum_{n=1}^{N-1}\left[\sigma_{n}^{+}\sigma_{n+1}^{-}+\sigma_{n}^{-}\sigma_{n+1}^{+}\right] \nonumber\\
& +  \frac{N^2}{8} +  N \alpha( \alpha -\frac{1}{2})   \\
& + \frac{1}{4}\sum_{n=1}^N \left[ n - N + (-1)^n (2\mu + \frac{1}{2}) - \frac{1}{2} \right] \sigma_{n}^{z} \nonumber \\
& +\sum_{n=1}^{N-1}\left(N-n\right) \left[ \alpha \sigma_{n}^{z}+\frac{1}{2} 
\underset{{\scriptstyle l<n}}{\sum}\sigma_{l}^{z}\sigma_{n}^{z} \right]. \nonumber
\label{obcham} 
\end{align}
In the continuum, the theory would be periodic under integer shifts of $\alpha$.  This is not the case on a finite lattice for OBC, but as we will see, it is the case for PBC.

The situation changes for PBC, where \eqref{Gauss}  leaves one quantum electric field degree of freedom $L \equiv L(0)$ unfixed, where $0$ is an arbitrarily chosen lattice site.   Eliminating all the electric field degrees of freedom except $ L $, we can write thes Hamiltonian for PBC as
\begin{align}
H_{\text{PBC}} & = x \sum_{n=1}^{N-1}\left[\sigma_{n}^{+}\sigma_{n+1}^{-}+\sigma_{n}^{-}\sigma_{n+1}^{+}\right] \nonumber \\
& +  x \left( \sigma_{N}^{+}e^{i\theta} \sigma_{1}^{-}+ \sigma_{N}^{-} e^{-i\theta} \sigma_{1}^{+} \right) \nonumber \\
& +\frac{N^{2}}{8}+N (L+\alpha)( L+ \alpha -\frac{1}{2})    \\
& + \frac{1}{4}\sum_{n=1}^N \left[ n - N + (-1)^n (2\mu + \frac{1}{2}) - \frac{1}{2} \right] \sigma_{n}^{z}   \nonumber \\
& +\sum_{n=1}^{N-1}\left(N-n\right) \left[ (L+\alpha) \sigma_{n}^{z}+\frac{1}{2} 
\underset{{\scriptstyle l<n}}{\sum}\sigma_{l}^{z}\sigma_{n}^{z} \right] . \nonumber
\label{hzham}
\end{align}
where  $ \theta \equiv \theta (0) $. The spectrum of $L$ is quantized in integer units by \eqref{lhopping} and is unbounded from above and below.   The theory is again specified by a real number $\alpha$, but shifts of $L$ by an integer can be absorbed by an opposite shift in $\alpha$, so one may restrict to the fundamental domain $0 \leq \alpha < 1$ or regard the theory as a periodic function of $\alpha$ (this is the well-known periodicity of the Schwinger model \cite{COLEMAN1975267}).  It is transitions between these states of $L$ that allow for the unwinding mechanism as discussed in \cite{kleban2011}, which cannot occur for OBC where $L$ is not a degree of freedom.

\section{Exact diagonalization} \label{sec:exact} 

To numerically solve the Schwinger model, we first compute the Hamiltonian matrix for $N$ lattice sites, using the basis defined by the eigenstates of $L$ and the $\sigma^z$ as described in the previous section. The physical subspace that we will study has total charge of zero, corresponding to total spin zero in the spin language $ \sum_n \sigma_n^z = 0 $.  In the case of PBC the zero charge condition is necessary for consistency with the boundary conditions (the total charge on a compact space must always vanish).  For OBC we will also impose that the total charge vanishes, which implies that the electric field is equal on the two boundaries.  One could consider non-zero total charge for OBC, which would correspond to an electric field gradient with different field values on the two boundaries, but we will not do so in this paper.  Restricting to  zero charge results in a  reduction in the Hilbert space dimension from $ 2^N $ to $  {N \choose N/2} \sim 2^N/\sqrt{N} $.  

 As explained in the previous section, there is an extra bosonic degree of freedom $L$ with PBC, so that the Hilbert space with PBC is  infinite dimensional even with $N$ finite.  However,  \eqref{hzham} shows that states with large values of $L$ have large energy.  Therefore for finite energy processes it is an accurate approximation to truncate the Hilbert space so that the magnitude of the electric field at any site $ n $ always falls within the range $-L_\text{max} \leq L_n \leq L_\text{max}$.   We verified that our results are insensitive to increasing $L_\text{max}$. The total dimension of the Hilbert space is then
\begin{align}
D_{PBC} = (2 L_\text{max}+1)D_\text{OBC} = (2 L_\text{max}+1){N \choose N/2}.
\end{align}
Due to the fast scaling with $ N $ we are restricted to relatively small numbers of lattice sites, but we will show that it is possible to see the relevant dynamics in such systems.  

To perform the time evolution, we first find the eigenstates $ | \epsilon_n \rangle  $ and eigenvalues $ \epsilon_n $ of the Hamiltonian matrix using a standard Python library. The time evolution is performed according to the relation 
\begin{align}
|\psi(t) \rangle & = e^{-iHt} |\psi(0) \rangle \nonumber \\
& = \sum_n e^{-i \epsilon_n t } \langle \epsilon_n |\psi(0) \rangle | \epsilon_n \rangle  ,
\label{timevolution}
\end{align}
where $ |\psi(0) \rangle $ is the initial state.  We will consider two types of  initial states.  The first corresponds to a quench, where the ground state is found without the presence of the background electric field
\begin{align}
 |\psi(0) \rangle = | \epsilon_0 (\alpha= 0 ) \rangle .
\label{initialcond1}
\end{align}
The initial state is then time evolved according to (\ref{timevolution}) with a non-zero value for the background field in the Hamiltonian.   The second type of initial state contains 
a pair introduced into the vacuum state,  given by
\begin{align}
|\psi(0) \rangle \propto \sigma^+_n \left( \prod_{j=n}^{m-1} e^{i \theta_j} \right) \sigma^-_{m} | \epsilon_0 \rangle.
\label{initialcond2}
\end{align}

This corresponds to an electron-positron pair located at sites $ n $ and $ m $ excited from the vacuum. The electric field shift operators $ e^{\pm i \theta_n} $ ensure that the resulting state satisfies Gauss' law.  The $\propto$ symbol reflects the fact that the state on the right-hand side of (\ref{initialcond2}) must be normalized.

In the case of PBC, when a pair is added  the field can be changed in two natural ways to satisfy \eqref{Gauss}.  Suppose a positron is inserted at site $n_p$ and an electron at site $n_e>n_p$.  Either the electric field is changed by $+g$ at all sites $n$ ``in between" the pair, $n_p \leq n < n_e$ as in \eqref{initialcond2}, or changed by $-g$ at all sites $n$ ``outside" the pair, i.e. for $n < n_p$ and $n \geq n_e$. We choose the initial quantum state that corresponds to an equal linear combination of these two possibilities.

We calculate several quantities of interest.  The expectation value of the charge density at site $n$ is
\begin{align}
\langle \rho_n(t) \rangle = {1 \over 2} \langle \psi(t) \left| \left( (-1)^n \sigma^z_n  + 1 \right) \right|   \psi(t) \rangle .
\label{cdense}
\end{align}
The expectation values of the  field $L_n = E(n)/g$ at site $n$, and the expectation value and standard deviation of the spatially averaged  field are
\begin{align} \label{Enfield} 
&  \langle L_n(t) \rangle = \, \langle \psi(t) | \left( L_n + \alpha \right)|  \psi(t) \rangle,    \nonumber \\  
& \langle L(t) \rangle= g^{-1} \langle E(t) \rangle =  N^{-1} \sum_{n=1}^N  \langle L_n(t) \rangle    \\
& \sigma_{L} = \sqrt{ \langle (L(t) - \langle L(t) \rangle )^{2} \rangle } . \nonumber
\end{align}
We also compute the probability to measure a specific field value value $l$ at site $n$ for the field $L_{n} = E(n)/g$ :
\begin{align} \label{Eprob} 
p(n, l) = \left| \langle \psi | n,l \rangle \right|^{2},
\end{align}
where $| n,l \rangle$ is the eigenstate of $L_{n}$ with eigenvalue $l$.

\section{Time evolution}
\label{sec:results}

\subsection{Massless limit}

The continuum Schwinger model is defined by the dimensionless parameters $\alpha$ and $m/g$.  The  limit $m/g \to 0$ describes massless charged fermions.  One might expect that the theory becomes singular in some way  because any non-zero electric field can be discharged immediately by the flow of massless charged particles.  Indeed,  QED in three spatial dimensions has a Landau pole at zero energy in this limit.  However the massless Schwinger model is not only regular, but Gaussian.  An initial non-zero electric field indeed discharges, but smoothly and with a finite frequency set by $g$.  

The continuum Schwinger model admits an exactly equivalent bosonized description, where the fermions $\psi$ and the electric field $E$ in \eqref{continuumham} are replaced by a single scalar field $\phi$ satisfying the relations $: \bar{\psi}\psi : = -c m \cos{(2\sqrt{\pi}\phi)}$,  $j^{\mu} = :\bar{\psi}\gamma^{\mu}\psi: = \pi^{-1/2}\epsilon^{\mu\nu}\partial_{\nu}\phi$, and
$F_{01} = e \pi^{-1/2}\phi$ (where $::$ denotes normal ordering) \cite{COLEMAN1975267}.
In the massless limit, the bosonized Hamiltonian takes the form
\begin{align}
{\cal H}_B & = {1 \over 2} \int dx \left( \dot \pi^2 + \left( {\partial \phi \over \partial x}\right)^2 + {g^2 \over \pi} (\phi - \alpha/\sqrt{\pi})^2 \right).
\label{bosonized}
\end{align}
With PBC and in the classical limit,  $\alpha \neq 0$ and the initial configuration $\phi = \dot \phi = 0$ would lead $\phi$ to oscillate sinusoidally in time and homogeneously in space, with period $2 \pi \sqrt{\pi}/g$.  On general grounds, the same should be true of $\langle \phi \rangle$, with the initial state the $\alpha=0$ ground state of the Hamiltonian (where $\langle \phi \rangle = 0$).  In fact, since the bosonized field theory is Gaussian, this initial state should evolve as a coherent state:  $\langle \phi \rangle$ should oscillate sinusoidally in time, with constant variance $\sigma_\phi^2 = \langle \left( \phi - \langle \phi \rangle \right)^2 \rangle$.  

\begin{figure}
\includegraphics[width=\columnwidth,scale = 2]{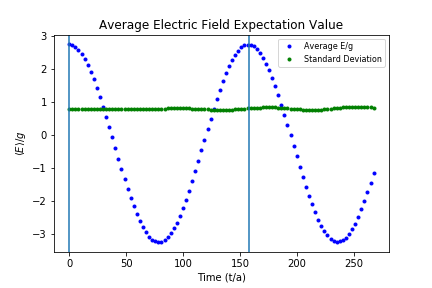}
\caption{Expectation value of the spatially averaged electric field  $\langle L \rangle  = \langle E \rangle /g $ in the massless limit, and its quantum standard deviation  (see \ref{Enfield}), with periodic boundary conditions (PBC) and parameters $m/g = 0, x=1/(ag)^2 = 200$. 
The initial state is described by (\ref{initialcond1}) with $\alpha=3$.
From the bosonized description (\ref{bosonized}) we expect  the expectation value of the field to oscillate sinusoidally with period $\Delta t/a = 2 \pi \sqrt{\pi x}$ and with constant standard deviation.  The vertical bars are separated by the analytic prediction for the period $\Delta t/a = 2 \pi \sqrt{\pi x}$. } 
\label{massless}
\end{figure}

The electric field in the original fermionic description is proportional to $\phi$, and so the prediciton  is that $\langle E(t) \rangle$ should oscillate sinusoidally in time with period $2 \pi \sqrt{\pi}/g$ and  constant variance.
 In Fig. \ref{massless} we plot the results of simulations in the massless regime that reproduce this behavior, including the correct period. The oscillations depicted in Fig. \ref{massless} can be thought of as a form of flux unwinding \cite{kleban2011}.  The initial value of the electric field is discharged by a current of positive charges flowing in one direction around the circle and negative charges flowing in the other.  The difference with the physics described in Fig. \ref{figloop} is that many charged pairs are involved -- the ground state of the massless theory contains a large density of charged pairs distributed homogeneously in space.  By contrast in the massive theory with large $m/g$ the ground state is close to the empty Fock space vacuum.   
  In the massless limit the gap between the first excited state and the ground state  can be computed analytically.  Using the bosonized description \eqref{bosonized}, one immediately obtains $\epsilon_1 - \epsilon_0 = g/\sqrt{\pi}$.  Our code accurately reproduces this value, as well as the spectrum found in \cite{byrnes2002density} for non-zero values of $m/g$.

\subsection{Time evolution of a charged pair in a background field}

\begin{figure}

\includegraphics[width=\columnwidth,scale = 2]{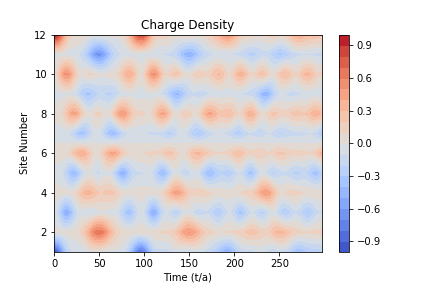}
\caption{Expectation value of the charge density for OBC (see \eqref{cdense}), with a pair inserted at $t=0$ at sites 1 and 12. Here we have $x=200$, $m/g=20$ and $\alpha=3$ which combined with the boundary condition gives a relatively high probability of transmission. The density has units of charge. }
\label{chargeplot}
\end{figure}

We now proceed to investigate the dynamics of the massive Schwinger model in the presence of an initial background field and in the massive regime $m/g \gg 1$.  The initial state we choose in these simulations is the ground state of the theory  with zero background field $\alpha = 0$, and with a charged pair added according to (\ref{initialcond2}).   We then evolve this state using the Hamiltonian with a non-zero value of $\alpha$, corresponding to turning on a background electric field. In the case of OBC, the value of the field on the boundaries is fixed to $\alpha$.  Semi-classically, when a pair is present in the initial state, we expect the field to accelerate the charges.  In Fig. \ref{chargeplot} we observe this behavior on a finite lattice by plotting the charge density as a function of time and lattice site.  Multiple bounces can be clearly seen, where the charges reflect off the boundary and primarily transmit through each other.  Because of the OBC, these multiple transmissions never reduce the field by more than two units (one for each charge).

\begin{figure}

\includegraphics[width=\columnwidth,scale = 2]{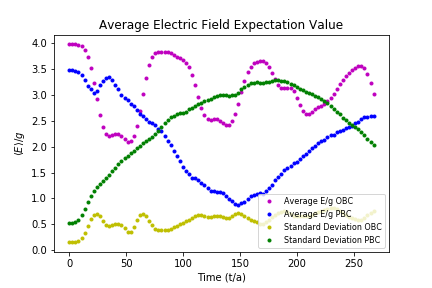}
\caption{Plot showing the expectation value and standard deviation of the spatially averaged electric field $\langle E \rangle /g $ (see \eqref{Enfield}) for OBC (cf.~Fig.~\ref{chargeplot}) and PBC (cf.~Figs.~\ref{probplotl3} and \ref{probplot}). For OBC,   $\langle E \rangle /g $ oscillates with an initial amplitude of roughly $2g$ as the charges bounce back and forth, while  the standard deviation of the field is roughly constant.  The oscillations damp because the transmission probability is less than one (see Fig.~\ref{chargeplot}).  For PBC, the larger decrease in the expectation value of the electric field  shows that particles can traverse  the circle multiple times, unwinding the initial field by multiple units.  The expectation value does not decrease all the way to minus its initial value because the transmission coefficient is not equal to one.  For the same reason the standard deviation of the electric field increases as the wavefunction spreads out in configuration space, with support on some configurations where  the field decreases to the maximum possible extent, and simultaneously on others where the field stays closer to its initial value.  Here $x=200$ and $m/g=20$, and a charged pair is inserted in the initial  state.}
\label{sttdev_increase}
\end{figure}

\begin{figure}

\includegraphics[width=\columnwidth,scale = 2]{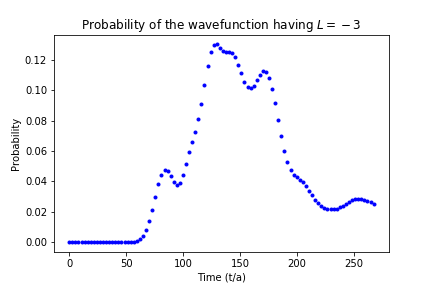}
\caption{The probability $p(n,l)$  (see \eqref{Eprob}) to measure an electric field $ l = E_n /g= -3$ at a specific lattice site $n$, as a function of time and with initial field $\langle E_n(t=0) \rangle/g \approx 3$. The parameters are the same as in Fig. \ref{sttdev_increase} above.  }
\label{probplotl3}
\end{figure}

\begin{figure}

\includegraphics[width=\columnwidth]{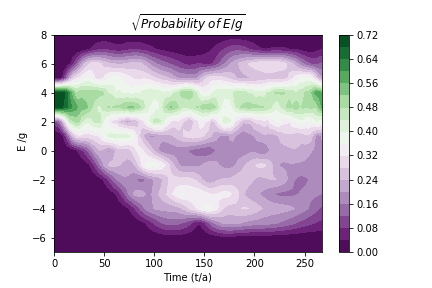}

\caption{The square root of the probability $\sqrt{p(n,l)}$ (see \eqref{Eprob}) to measure an electric field $ l = E_n /g $ at a specific lattice site $n$,  as a function of time and $ l$.  The square root is taken for ease of visualization.    The parameters are identical to those of Fig.~\ref{probplotl3} and Fig.~\ref{sttdev_increase};  Fig.~\ref{probplotl3}  corresponds to a horizontal slice of this figure at $E/g = -3$.  One can see that the initial expectation value of the field $\langle E_n \rangle/g \approx 3.5 $ remains quite probable for all times, but more and more negative values, down to roughly $ E_n /g=-3$, attain non-negligible probability as time passes.  At later times this behavior approximately reverses as the field rewinds.  This is  in accord with the analytic predictions of flux unwinding. \label{probplot}}
\end{figure}

In the case of PBC, as illustrated in Fig.~\ref{figloop},  multiple units of the initial flux can unwind due to a  charged pair traversing the circle multiple times \cite{kleban2011}. Starting from a background field of $ \alpha $, each pass of a charge around the circle removes an additional unit of flux, and so the field should decrease steadily in time until the background electric field becomes $ -\alpha $. Conservation of energy dictates the charges should come to rest at this point and there can be no further decrease in the field (see Fig. \ref{probplot}).   This will be followed by a phase of ``rewinding" where the process occurs in reverse.

In the massive theory and in the regime of parameters we can reach with our simulations, the transmission probability is never extremely close to one.  Hence after multiple would-be transmissions the wavefunction will evolve into a broad class of configurations.  In some configurations the field has unwound to the maximum extent possible due to an unbroken chain of transmissions.  In others, some reflections have occurred (or possibly transitions to other states), preventing the field from unwinding.  In contrast to the massless regime, this should lead to an increase in the standard deviation in $L$ for some time, and the expectation value of the field will not decrease all the way to minus its initial value (see Fig. \ref{sttdev_increase}). 

Nevertheless, even in this fully quantum regime, certain features provide a clear signal that unwinding is occurring.   One is the behavior of probability $p( L, t)$ (derived from the wavefunction) to measure an electric field value $L$ at time $t$.  According to the semi-classical analysis of \cite{kleban2011} in a regime where transmission of two opposite charges through each other is highly probable, the maximum possible unwinding  corresponds to $L = - \alpha $ (because then the field $L$ has changed from $\alpha$ to $-\alpha$).  

 The quantum mechanical probability $p( L, t)$ should behave in a way that corresponds to this semi-classical physics.  That is, for  $ L$ in the range $-\alpha < L< \alpha$, $p( L, t)$  should increase monotonically with $t$ until a certain time when the unwinding has reached its apex, and decrease with decreasing $ L$ for any fixed $t$ during this time.  For later times, this behavior should (roughly) reverse.  Furthermore $p( L, t)$ should be very small for $ L < - \alpha$ and $ L > \alpha$ for all times, as these regimes are classically forbidden by conservation of energy.  In Figs.~\ref{probplotl3} and \ref{probplot} we plot $p( L, t)$, which indeed agrees with these expectations.

\subsection{Flux unwinding by Schwinger pair production}

\begin{figure}

\includegraphics[width=\columnwidth]{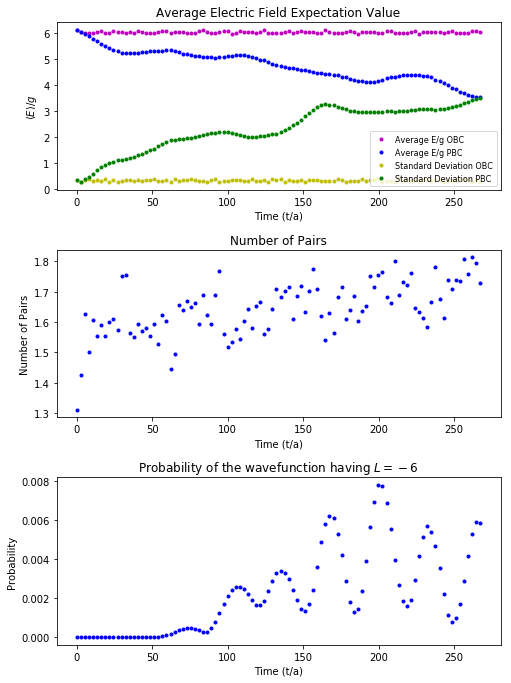}

\caption{Simulations where the initial state is the zero-electric field ground state, time-evolved with an applied background field $\alpha = 6$. The parameters are $x = 50$,  $m/g=2$, and $N=10$. Top pane: the expectation value and standard deviation in the spatially averaged electric field for both PBC and OBC, as in Fig.~\ref{sttdev_increase}  (see \eqref{Enfield}).  One can see that unwinding is occurring by the decrease in the expectation value and the increase in the standard deviation. Middle pane:  the number of pairs as a function of time for PBC.  The initial value is non-zero because for $m/g = 2$ the ground state does not coincide with the Fock space vacuum.  Nevertheless the increase shows that Schwinger pair production is occurring.   Bottom pane: the probability of measuring $L = -6$ at a specific lattice site for PBC, similar to  Fig.~\ref{probplotl3}.  As expected from unwinding, the probability is essentially zero until enough time has passed for a produced pair to propagate around the circle.   The  probability is suppressed compared  to Fig.~\ref{probplotl3}, because to reach $L=-6$ a pair must first be produced and then traverse the circle twice as many times.}
\label{ppfig}
\end{figure}

In this section we take the initial state to be the ground state with $\alpha=0$, and then time-evolve it using the Hamiltonian \eqref{obcham} or \eqref{hzham} with $\alpha \neq 0$.  In some parameter regimes we expect Schwinger pair-production to occur.  Since pair production conserves energy, the mass-energy of the charged pair $2 m$ must be balanced by the change in energy density due to the reduction in the background field in between the charges.  If $E_i $ is the initial  background field and the charges are at rest and separated by a distance $d$, conservation of energy requires
$$
 2 m  = -{1 \over 2 } \int \left( E_f^2 - E_i^2 \right) dy = d    \left( g E_i  -{g^2 \over 2} \right),
 $$
or in our notation $L = E/g, \, x = 1/(ag)^2$,
$$
{d \over a} = 2 \sqrt{x} \left({m \over g}\right)  {1 \over L_i - 1/2}\, \,.
$$
If $d/a > N$ (where $N$ is the number of lattice sites) there is not enough room on the interval for a pair to be produced (although with PBC the electric field can still decay with time, but it will do so by tunneling homogeneously on the circle, not by pair production).  

To study the effects of Schwinger pair production we  consider $x = 50$, $\langle L_n \rangle = 6$,  $m/g=2$, and $N=10$.  With these values $d/a = 5.1 < N$ and we expect pair production -- followed by unwinding for PBC -- to take place.  However in contrast to the previous section where a massive charged pair was inserted by hand, the probability of unwinding will be substantially reduced because the rate of pair production per time per length in the initial state $\Gamma_\text{pp}$ is exponentially suppressed.  In the continuum theory, 
$$
a^2 \Gamma_\text{pp} \sim {L \over x} \exp \left[- \pi {\left(m \over g\right)}^2 { 1 \over L - 1/2} \right] \approx .01,
$$
where the  $\approx$ holds with the values listed above.

For OBC, even when $d/a < N$ such that a pair can be produced, the field cannot be reduced significantly because the pairs simply bounce back and forth inside the interval.  Furthermore the lack of dissipation prevents charges from accumulating near the boundaries.  This together with the low rate of pair production and the spatial homogeneity  of the quantum state results in nearly no change in the expectation value or standard deviation of the spatially averaged field.   The results  for both PBC and OBC are plotted in Fig.~\ref{ppfig}.

\section{Conclusion}

\label{sec:conc}

We have performed a finite lattice simulation of the dynamics of the massive Schwinger model.
Starting from the ground state of the finite lattice Hamiltonian with zero electric field, we apply an electric field and observe its behavior as a function of time.  When the parameter $m/g$ is small, the field expectation value oscillates sinusoidally with constant standard deviation as expected from the bosonized description of the theory.  We also considered the effect of directly introducing charges to the initial state, and observing their dynamics within an  applied field.  When $m/g$ is large, our results are consistent with the semi-classical picture of a pair being accelerated by the field and unwinding the  background field.  
We verified that the charges can transmit through each other with an $\mathcal{O}(1)$ probability, again consistent with the expectation from a semi-classical analysis.  Lastly, we verified that flux unwinding can also occur due to Schwinger pair production. 

In the future we plan to numerically investigate the Schwinger model in an expanding, de Sitter background spacetime.  This will allow us to study the ``hyperconductivity" phenomenon proposed in \cite{Frob:2014zka}, as well as the possibility of using the unwinding dynamics to drive inflation \cite{damico2012}.

\begin{acknowledgments}
T.~B.~is supported by the Shanghai Research Challenge Fund; New York University Global Seed Grants for Collaborative Research; National Natural Science Foundation of China (61571301); the Thousand Talents Program for Distinguished Young Scholars (D1210036A); and the NSFC Research Fund for International Young Scientists (11650110425); NYU-ECNU Institute of Physics at NYU Shanghai; the Science and Technology Commission of Shanghai Municipality (17ZR1443600); the China Science and Technology Exchange Center (NGA-16-001); and the NSFC-RFBR Collaborative grant (81811530112). The work of MK is supported in part by the NSF through grants PHY-1214302 and PHY-1820814, and he acknowledges membership at the NYU- ECNU Joint Physics Research Institute in Shanghai.
\end{acknowledgments}

\bibliography{bibliography}

\begin{thebibliography}{21}%
\makeatletter
\providecommand \@ifxundefined [1]{%
 \@ifx{#1\undefined}
}%
\providecommand \@ifnum [1]{%
 \ifnum #1\expandafter \@firstoftwo
 \else \expandafter \@secondoftwo
 \fi
}%
\providecommand \@ifx [1]{%
 \ifx #1\expandafter \@firstoftwo
 \else \expandafter \@secondoftwo
 \fi
}%
\providecommand \natexlab [1]{#1}%
\providecommand \enquote  [1]{``#1''}%
\providecommand \bibnamefont  [1]{#1}%
\providecommand \bibfnamefont [1]{#1}%
\providecommand \citenamefont [1]{#1}%
\providecommand \href@noop [0]{\@secondoftwo}%
\providecommand \href [0]{\begingroup \@sanitize@url \@href}%
\providecommand \@href[1]{\@@startlink{#1}\@@href}%
\providecommand \@@href[1]{\endgroup#1\@@endlink}%
\providecommand \@sanitize@url [0]{\catcode `\\12\catcode `\$12\catcode
  `\&12\catcode `\#12\catcode `\^12\catcode `\_12\catcode `\%12\relax}%
\providecommand \@@startlink[1]{}%
\providecommand \@@endlink[0]{}%
\providecommand \url  [0]{\begingroup\@sanitize@url \@url }%
\providecommand \@url [1]{\endgroup\@href {#1}{\urlprefix }}%
\providecommand \urlprefix  [0]{URL }%
\providecommand \Eprint [0]{\href }%
\providecommand \doibase [0]{http://dx.doi.org/}%
\providecommand \selectlanguage [0]{\@gobble}%
\providecommand \bibinfo  [0]{\@secondoftwo}%
\providecommand \bibfield  [0]{\@secondoftwo}%
\providecommand \translation [1]{[#1]}%
\providecommand \BibitemOpen [0]{}%
\providecommand \bibitemStop [0]{}%
\providecommand \bibitemNoStop [0]{.\EOS\space}%
\providecommand \EOS [0]{\spacefactor3000\relax}%
\providecommand \BibitemShut  [1]{\csname bibitem#1\endcsname}%
\let\auto@bib@innerbib\@empty
\bibitem [{\citenamefont {Schwinger}(1962)}]{schwinger1962gauge}%
  \BibitemOpen
  \bibfield  {author} {\bibinfo {author} {\bibfnamefont {J.}~\bibnamefont
  {Schwinger}},\ }\bibfield  {title} {\enquote {\bibinfo {title} {Gauge
  invariance and mass. ii},}\ }\href@noop {} {\bibfield  {journal} {\bibinfo
  {journal} {Physical Review}\ }\textbf {\bibinfo {volume} {128}},\ \bibinfo
  {pages} {2425} (\bibinfo {year} {1962})}\BibitemShut {NoStop}%
\bibitem [{\citenamefont {Coleman}\ \emph {et~al.}(1975)\citenamefont
  {Coleman}, \citenamefont {Jackiw},\ and\ \citenamefont
  {Susskind}}]{COLEMAN1975267}%
  \BibitemOpen
  \bibfield  {author} {\bibinfo {author} {\bibfnamefont {S.}~\bibnamefont
  {Coleman}}, \bibinfo {author} {\bibfnamefont {R.}~\bibnamefont {Jackiw}}, \
  and\ \bibinfo {author} {\bibfnamefont {L.}~\bibnamefont {Susskind}},\
  }\bibfield  {title} {\enquote {\bibinfo {title} {Charge shielding and quark
  confinement in the massive schwinger model},}\ }\href {\doibase
  https://doi.org/10.1016/0003-4916(75)90212-2} {\bibfield  {journal} {\bibinfo
   {journal} {Annals of Physics}\ }\textbf {\bibinfo {volume} {93}},\ \bibinfo
  {pages} {267 -- 275} (\bibinfo {year} {1975})}\BibitemShut {NoStop}%
\bibitem [{\citenamefont {Coleman}(1976)}]{Coleman:1976uz}%
  \BibitemOpen
  \bibfield  {author} {\bibinfo {author} {\bibfnamefont {S.}~\bibnamefont
  {Coleman}},\ }\bibfield  {title} {\enquote {\bibinfo {title} {{More About the
  Massive Schwinger Model}},}\ }\href {\doibase 10.1016/0003-4916(76)90280-3}
  {\bibfield  {journal} {\bibinfo  {journal} {Annals Phys.}\ }\textbf {\bibinfo
  {volume} {101}},\ \bibinfo {pages} {239} (\bibinfo {year}
  {1976})}\BibitemShut {NoStop}%
\bibitem [{\citenamefont {D'Amico}\ \emph {et~al.}(2013)\citenamefont
  {D'Amico}, \citenamefont {Gobbetti}, \citenamefont {Kleban},\ and\
  \citenamefont {Schillo}}]{damico2012}%
  \BibitemOpen
  \bibfield  {author} {\bibinfo {author} {\bibfnamefont {G.}~\bibnamefont
  {D'Amico}}, \bibinfo {author} {\bibfnamefont {R.}~\bibnamefont {Gobbetti}},
  \bibinfo {author} {\bibfnamefont {M.}~\bibnamefont {Kleban}}, \ and\ \bibinfo
  {author} {\bibfnamefont {Marjorie}\ \bibnamefont {Schillo}},\ }\bibfield
  {title} {\enquote {\bibinfo {title} {{Unwinding Inflation}},}\ }\href
  {\doibase 10.1088/1475-7516/2013/03/004} {\bibfield  {journal} {\bibinfo
  {journal} {JCAP}\ }\textbf {\bibinfo {volume} {1303}},\ \bibinfo {pages}
  {004} (\bibinfo {year} {2013})},\ \Eprint {http://arxiv.org/abs/1211.4589}
  {arXiv:1211.4589 [hep-th]} \BibitemShut {NoStop}%
\bibitem [{\citenamefont {Hebenstreit}\ \emph {et~al.}(2013)\citenamefont
  {Hebenstreit}, \citenamefont {Berges},\ and\ \citenamefont
  {Gelfand}}]{hebenstreit2013}%
  \BibitemOpen
  \bibfield  {author} {\bibinfo {author} {\bibfnamefont {F.}~\bibnamefont
  {Hebenstreit}}, \bibinfo {author} {\bibfnamefont {J.}~\bibnamefont {Berges}},
  \ and\ \bibinfo {author} {\bibfnamefont {D.}~\bibnamefont {Gelfand}},\
  }\bibfield  {title} {\enquote {\bibinfo {title} {{Real-time dynamics of
  string breaking}},}\ }\href {\doibase 10.1103/PhysRevLett.111.201601}
  {\bibfield  {journal} {\bibinfo  {journal} {Phys. Rev. Lett.}\ }\textbf
  {\bibinfo {volume} {111}},\ \bibinfo {pages} {201601} (\bibinfo {year}
  {2013})},\ \Eprint {http://arxiv.org/abs/1307.4619} {arXiv:1307.4619
  [hep-ph]} \BibitemShut {NoStop}%
\bibitem [{\citenamefont {Buyens}\ \emph {et~al.}(2017)\citenamefont {Buyens},
  \citenamefont {Haegeman}, \citenamefont {Hebenstreit}, \citenamefont
  {Verstraete},\ and\ \citenamefont {Van~Acoleyen}}]{buyens2016}%
  \BibitemOpen
  \bibfield  {author} {\bibinfo {author} {\bibfnamefont {B.}~\bibnamefont
  {Buyens}}, \bibinfo {author} {\bibfnamefont {J.}~\bibnamefont {Haegeman}},
  \bibinfo {author} {\bibfnamefont {F.}~\bibnamefont {Hebenstreit}}, \bibinfo
  {author} {\bibfnamefont {Frank}\ \bibnamefont {Verstraete}}, \ and\ \bibinfo
  {author} {\bibfnamefont {Karel}\ \bibnamefont {Van~Acoleyen}},\ }\bibfield
  {title} {\enquote {\bibinfo {title} {{Real-time simulation of the Schwinger
  effect with Matrix Product States}},}\ }\href {\doibase
  10.1103/PhysRevD.96.114501} {\bibfield  {journal} {\bibinfo  {journal} {Phys.
  Rev.}\ }\textbf {\bibinfo {volume} {D96}},\ \bibinfo {pages} {114501}
  (\bibinfo {year} {2017})},\ \Eprint {http://arxiv.org/abs/1612.00739}
  {arXiv:1612.00739 [hep-lat]} \BibitemShut {NoStop}%
\bibitem [{\citenamefont {Heisenberg}\ and\ \citenamefont
  {Euler}(1936)}]{heisenberg1935}%
  \BibitemOpen
  \bibfield  {author} {\bibinfo {author} {\bibfnamefont {W.}~\bibnamefont
  {Heisenberg}}\ and\ \bibinfo {author} {\bibfnamefont {H.}~\bibnamefont
  {Euler}},\ }\bibfield  {title} {\enquote {\bibinfo {title} {{Folgerungen aus
  der Diracschen Theorie des Positrons}},}\ }\href {\doibase
  10.1007/BF01343663} {\bibfield  {journal} {\bibinfo  {journal} {Z. Phys.}\
  }\textbf {\bibinfo {volume} {98}},\ \bibinfo {pages} {714--732} (\bibinfo
  {year} {1936})},\ \Eprint {http://arxiv.org/abs/physics/0605038}
  {arXiv:physics/0605038 [physics]} \BibitemShut {NoStop}%
\bibitem [{\citenamefont {Kleban}\ \emph {et~al.}(2011)\citenamefont {Kleban},
  \citenamefont {Krishnaiyengar},\ and\ \citenamefont {Porrati}}]{kleban2011}%
  \BibitemOpen
  \bibfield  {author} {\bibinfo {author} {\bibfnamefont {M.}~\bibnamefont
  {Kleban}}, \bibinfo {author} {\bibfnamefont {K.}~\bibnamefont
  {Krishnaiyengar}}, \ and\ \bibinfo {author} {\bibfnamefont {M.}~\bibnamefont
  {Porrati}},\ }\bibfield  {title} {\enquote {\bibinfo {title} {{Flux Discharge
  Cascades in Various Dimensions}},}\ }\href {\doibase 10.1007/JHEP11(2011)096}
  {\bibfield  {journal} {\bibinfo  {journal} {JHEP}\ }\textbf {\bibinfo
  {volume} {11}},\ \bibinfo {pages} {096} (\bibinfo {year} {2011})},\ \Eprint
  {http://arxiv.org/abs/1108.6102} {arXiv:1108.6102 [hep-th]} \BibitemShut
  {NoStop}%
\bibitem [{\citenamefont {{Silverstein}}\ and\ \citenamefont
  {{Westphal}}(2008)}]{2008PhRvD..78j6003S}%
  \BibitemOpen
  \bibfield  {author} {\bibinfo {author} {\bibfnamefont {E.}~\bibnamefont
  {{Silverstein}}}\ and\ \bibinfo {author} {\bibfnamefont {A.}~\bibnamefont
  {{Westphal}}},\ }\bibfield  {title} {\enquote {\bibinfo {title} {{Monodromy
  in the CMB: Gravity waves and string inflation}},}\ }\href {\doibase
  10.1103/PhysRevD.78.106003} {\bibfield  {journal} {\bibinfo  {journal}
  {\prd}\ }\textbf {\bibinfo {volume} {78}},\ \bibinfo {eid} {106003} (\bibinfo
  {year} {2008})},\ \Eprint {http://arxiv.org/abs/0803.3085} {arXiv:0803.3085
  [hep-th]} \BibitemShut {NoStop}%
\bibitem [{\citenamefont {{McAllister}}\ \emph {et~al.}(2010)\citenamefont
  {{McAllister}}, \citenamefont {{Silverstein}},\ and\ \citenamefont
  {{Westphal}}}]{2010PhRvD..82d6003M}%
  \BibitemOpen
  \bibfield  {author} {\bibinfo {author} {\bibfnamefont {L.}~\bibnamefont
  {{McAllister}}}, \bibinfo {author} {\bibfnamefont {E.}~\bibnamefont
  {{Silverstein}}}, \ and\ \bibinfo {author} {\bibfnamefont {A.}~\bibnamefont
  {{Westphal}}},\ }\bibfield  {title} {\enquote {\bibinfo {title} {{Gravity
  waves and linear inflation from axion monodromy}},}\ }\href {\doibase
  10.1103/PhysRevD.82.046003} {\bibfield  {journal} {\bibinfo  {journal}
  {\prd}\ }\textbf {\bibinfo {volume} {82}},\ \bibinfo {eid} {046003} (\bibinfo
  {year} {2010})},\ \Eprint {http://arxiv.org/abs/0808.0706} {arXiv:0808.0706
  [hep-th]} \BibitemShut {NoStop}%
\bibitem [{\citenamefont {{D'Amico}}\ \emph {et~al.}(2013)\citenamefont
  {{D'Amico}}, \citenamefont {{Gobbetti}}, \citenamefont {{Kleban}},\ and\
  \citenamefont {{Schillo}}}]{2013PhLB..725..218D}%
  \BibitemOpen
  \bibfield  {author} {\bibinfo {author} {\bibfnamefont {G.}~\bibnamefont
  {{D'Amico}}}, \bibinfo {author} {\bibfnamefont {R.}~\bibnamefont
  {{Gobbetti}}}, \bibinfo {author} {\bibfnamefont {M.}~\bibnamefont
  {{Kleban}}}, \ and\ \bibinfo {author} {\bibfnamefont {M.~L.}\ \bibnamefont
  {{Schillo}}},\ }\bibfield  {title} {\enquote {\bibinfo {title} {{Inflation
  from flux cascades}},}\ }\href {\doibase 10.1016/j.physletb.2013.07.050}
  {\bibfield  {journal} {\bibinfo  {journal} {Physics Letters B}\ }\textbf
  {\bibinfo {volume} {725}},\ \bibinfo {pages} {218--222} (\bibinfo {year}
  {2013})},\ \Eprint {http://arxiv.org/abs/1211.3416} {arXiv:1211.3416
  [hep-th]} \BibitemShut {NoStop}%
\bibitem [{\citenamefont {Ba{\~n}uls}\ \emph {et~al.}(2013)\citenamefont
  {Ba{\~n}uls}, \citenamefont {Cichy}, \citenamefont {Cirac}, \citenamefont
  {Jansen},\ and\ \citenamefont {Saito}}]{banuls2013matrix}%
  \BibitemOpen
  \bibfield  {author} {\bibinfo {author} {\bibfnamefont {M.~C.}\ \bibnamefont
  {Ba{\~n}uls}}, \bibinfo {author} {\bibfnamefont {K.}~\bibnamefont {Cichy}},
  \bibinfo {author} {\bibfnamefont {J.~I.}\ \bibnamefont {Cirac}}, \bibinfo
  {author} {\bibfnamefont {K.}~\bibnamefont {Jansen}}, \ and\ \bibinfo {author}
  {\bibfnamefont {H.}~\bibnamefont {Saito}},\ }\bibfield  {title} {\enquote
  {\bibinfo {title} {Matrix product states for lattice field theories},}\
  }\href@noop {} {\bibfield  {journal} {\bibinfo  {journal} {arXiv preprint
  arXiv:1310.4118}\ } (\bibinfo {year} {2013})}\BibitemShut {NoStop}%
\bibitem [{\citenamefont {Ba{\~n}uls}\ \emph {et~al.}(2015)\citenamefont
  {Ba{\~n}uls}, \citenamefont {Cichy}, \citenamefont {Cirac}, \citenamefont
  {Jansen},\ and\ \citenamefont {Saito}}]{banuls2015thermal}%
  \BibitemOpen
  \bibfield  {author} {\bibinfo {author} {\bibfnamefont {M.~C.}\ \bibnamefont
  {Ba{\~n}uls}}, \bibinfo {author} {\bibfnamefont {K.}~\bibnamefont {Cichy}},
  \bibinfo {author} {\bibfnamefont {J.~I.}\ \bibnamefont {Cirac}}, \bibinfo
  {author} {\bibfnamefont {K.}~\bibnamefont {Jansen}}, \ and\ \bibinfo {author}
  {\bibfnamefont {H.}~\bibnamefont {Saito}},\ }\bibfield  {title} {\enquote
  {\bibinfo {title} {Thermal evolution of the schwinger model with matrix
  product operators},}\ }\href@noop {} {\bibfield  {journal} {\bibinfo
  {journal} {Physical Review D}\ }\textbf {\bibinfo {volume} {92}},\ \bibinfo
  {pages} {034519} (\bibinfo {year} {2015})}\BibitemShut {NoStop}%
\bibitem [{\citenamefont {Ba{\~n}uls}\ \emph {et~al.}(2017)\citenamefont
  {Ba{\~n}uls}, \citenamefont {Cichy}, \citenamefont {Cirac}, \citenamefont
  {Jansen},\ and\ \citenamefont {K{\"u}hn}}]{banuls2017density}%
  \BibitemOpen
  \bibfield  {author} {\bibinfo {author} {\bibfnamefont {M.~C.}\ \bibnamefont
  {Ba{\~n}uls}}, \bibinfo {author} {\bibfnamefont {K.}~\bibnamefont {Cichy}},
  \bibinfo {author} {\bibfnamefont {J.~I.}\ \bibnamefont {Cirac}}, \bibinfo
  {author} {\bibfnamefont {K.}~\bibnamefont {Jansen}}, \ and\ \bibinfo {author}
  {\bibfnamefont {S.}~\bibnamefont {K{\"u}hn}},\ }\bibfield  {title} {\enquote
  {\bibinfo {title} {Density induced phase transitions in the schwinger model:
  A study with matrix product states},}\ }\href@noop {} {\bibfield  {journal}
  {\bibinfo  {journal} {Physical Review Letters}\ }\textbf {\bibinfo {volume}
  {118}},\ \bibinfo {pages} {071601} (\bibinfo {year} {2017})}\BibitemShut
  {NoStop}%
\bibitem [{\citenamefont {Manton}(1985)}]{manton1985}%
  \BibitemOpen
  \bibfield  {author} {\bibinfo {author} {\bibfnamefont {N.~S.}\ \bibnamefont
  {Manton}},\ }\bibfield  {title} {\enquote {\bibinfo {title} {{The Schwinger
  Model and Its Axial Anomaly}},}\ }\href {\doibase
  10.1016/0003-4916(85)90199-X} {\bibfield  {journal} {\bibinfo  {journal}
  {Annals Phys.}\ }\textbf {\bibinfo {volume} {159}},\ \bibinfo {pages}
  {220--251} (\bibinfo {year} {1985})}\BibitemShut {NoStop}%
\bibitem [{\citenamefont {Lowenstein}\ and\ \citenamefont
  {Swieca}(1971)}]{lowenstein1971quantum}%
  \BibitemOpen
  \bibfield  {author} {\bibinfo {author} {\bibfnamefont {J.~H.}\ \bibnamefont
  {Lowenstein}}\ and\ \bibinfo {author} {\bibfnamefont {J.~A.}\ \bibnamefont
  {Swieca}},\ }\bibfield  {title} {\enquote {\bibinfo {title} {Quantum
  electrodynamics in two dimensions},}\ }\href@noop {} {\bibfield  {journal}
  {\bibinfo  {journal} {Annals of Physics}\ }\textbf {\bibinfo {volume} {68}},\
  \bibinfo {pages} {172--195} (\bibinfo {year} {1971})}\BibitemShut {NoStop}%
\bibitem [{\citenamefont {Byrnes}\ \emph {et~al.}(2002)\citenamefont {Byrnes},
  \citenamefont {Sriganesh}, \citenamefont {Bursill},\ and\ \citenamefont
  {Hamer}}]{byrnes2002density}%
  \BibitemOpen
  \bibfield  {author} {\bibinfo {author} {\bibfnamefont {T.~M.~R.}\
  \bibnamefont {Byrnes}}, \bibinfo {author} {\bibfnamefont {P.}~\bibnamefont
  {Sriganesh}}, \bibinfo {author} {\bibfnamefont {R.~J.}\ \bibnamefont
  {Bursill}}, \ and\ \bibinfo {author} {\bibfnamefont {C.~J.}\ \bibnamefont
  {Hamer}},\ }\bibfield  {title} {\enquote {\bibinfo {title} {Density matrix
  renormalization group approach to the massive schwinger model},}\ }\href@noop
  {} {\bibfield  {journal} {\bibinfo  {journal} {Physical Review D}\ }\textbf
  {\bibinfo {volume} {66}},\ \bibinfo {pages} {013002} (\bibinfo {year}
  {2002})}\BibitemShut {NoStop}%
\bibitem [{\citenamefont {Kogut}\ and\ \citenamefont
  {Susskind}(1975)}]{kogut1975hamiltonian}%
  \BibitemOpen
  \bibfield  {author} {\bibinfo {author} {\bibfnamefont {J.}~\bibnamefont
  {Kogut}}\ and\ \bibinfo {author} {\bibfnamefont {L.}~\bibnamefont
  {Susskind}},\ }\bibfield  {title} {\enquote {\bibinfo {title} {Hamiltonian
  formulation of wilson's lattice gauge theories},}\ }\href@noop {} {\bibfield
  {journal} {\bibinfo  {journal} {Physical Review D}\ }\textbf {\bibinfo
  {volume} {11}},\ \bibinfo {pages} {395} (\bibinfo {year} {1975})}\BibitemShut
  {NoStop}%
\bibitem [{\citenamefont {Carroll}\ \emph {et~al.}(1976)\citenamefont
  {Carroll}, \citenamefont {Kogut}, \citenamefont {Sinclair},\ and\
  \citenamefont {Susskind}}]{carroll1976lattice}%
  \BibitemOpen
  \bibfield  {author} {\bibinfo {author} {\bibfnamefont {A.}~\bibnamefont
  {Carroll}}, \bibinfo {author} {\bibfnamefont {J.}~\bibnamefont {Kogut}},
  \bibinfo {author} {\bibfnamefont {D.~K.}\ \bibnamefont {Sinclair}}, \ and\
  \bibinfo {author} {\bibfnamefont {L.}~\bibnamefont {Susskind}},\ }\bibfield
  {title} {\enquote {\bibinfo {title} {Lattice gauge theory calculations in 1+
  1 dimensions and the approach to the continuum limit},}\ }\href@noop {}
  {\bibfield  {journal} {\bibinfo  {journal} {Physical Review D}\ }\textbf
  {\bibinfo {volume} {13}},\ \bibinfo {pages} {2270} (\bibinfo {year}
  {1976})}\BibitemShut {NoStop}%
\bibitem [{\citenamefont {Banks}\ \emph {et~al.}(1976)\citenamefont {Banks},
  \citenamefont {Susskind},\ and\ \citenamefont {Kogut}}]{banks1976strong}%
  \BibitemOpen
  \bibfield  {author} {\bibinfo {author} {\bibfnamefont {T.}~\bibnamefont
  {Banks}}, \bibinfo {author} {\bibfnamefont {L.}~\bibnamefont {Susskind}}, \
  and\ \bibinfo {author} {\bibfnamefont {J.}~\bibnamefont {Kogut}},\ }\bibfield
   {title} {\enquote {\bibinfo {title} {Strong-coupling calculations of lattice
  gauge theories:(1+ 1)-dimensional exercises},}\ }\href@noop {} {\bibfield
  {journal} {\bibinfo  {journal} {Physical Review D}\ }\textbf {\bibinfo
  {volume} {13}},\ \bibinfo {pages} {1043} (\bibinfo {year}
  {1976})}\BibitemShut {NoStop}%
\bibitem [{\citenamefont {Fröb}\ \emph {et~al.}(2014)\citenamefont {Fröb},
  \citenamefont {Garriga}, \citenamefont {Kanno}, \citenamefont {Sasaki},
  \citenamefont {Soda}, \citenamefont {Tanaka},\ and\ \citenamefont
  {Vilenkin}}]{Frob:2014zka}%
  \BibitemOpen
  \bibfield  {author} {\bibinfo {author} {\bibfnamefont {M.~B.}\ \bibnamefont
  {Fröb}}, \bibinfo {author} {\bibfnamefont {J.}~\bibnamefont {Garriga}},
  \bibinfo {author} {\bibfnamefont {S.}~\bibnamefont {Kanno}}, \bibinfo
  {author} {\bibfnamefont {Misao}\ \bibnamefont {Sasaki}}, \bibinfo {author}
  {\bibfnamefont {Jiro}\ \bibnamefont {Soda}}, \bibinfo {author} {\bibfnamefont
  {Takahiro}\ \bibnamefont {Tanaka}}, \ and\ \bibinfo {author} {\bibfnamefont
  {Alexander}\ \bibnamefont {Vilenkin}},\ }\bibfield  {title} {\enquote
  {\bibinfo {title} {{Schwinger effect in de Sitter space}},}\ }\href {\doibase
  10.1088/1475-7516/2014/04/009} {\bibfield  {journal} {\bibinfo  {journal}
  {JCAP}\ }\textbf {\bibinfo {volume} {1404}},\ \bibinfo {pages} {009}
  (\bibinfo {year} {2014})},\ \Eprint {http://arxiv.org/abs/1401.4137}
  {arXiv:1401.4137 [hep-th]} \BibitemShut {NoStop}%
\end{thebibliography}%

\end{document}